\documentclass[prb,superscriptaddress,twocolumn,longbibliography]{revtex4-1}
\usepackage{times,amsmath}
\usepackage{epsfig}
\usepackage{color}
\usepackage{longtable}
\usepackage{ulem}
\usepackage{graphicx}
\usepackage{dcolumn}
\usepackage{bm}
\usepackage{bookmark}
\usepackage{tabularx}
\usepackage{hyperref}
\usepackage{multirow}

\hypersetup{colorlinks=true, citecolor=blue, filecolor=blue, linkcolor=blue, urlcolor=blue}

\begin{document}

\preprint{APS/123-QED}

\title{Spectral Neural Network Potentials for Binary Alloys}

\author{David Zagaceta}
\affiliation{Department of Physics and Astronomy, University of Nevada, Las Vegas, NV 89154, USA}

\author{Howard Yanxon}
\affiliation{Department of Physics and Astronomy, University of Nevada, Las Vegas, NV 89154, USA}

\author{Qiang Zhu}
\email{qiang.zhu@unlv.edu}
\affiliation{Department of Physics and Astronomy, University of Nevada, Las Vegas, NV 89154, USA}

\date{\today}

\begin{abstract}
In this work, we present a numerical implementation to compute the atom centered descriptors introduced by Bartok et al (Phys. Rev. B, 87, 184115, 2013) based on the harmonic analysis of the atomic neighbor density function.  
Specifically, we focus on two types of descriptors, the smooth SO(3) power spectrum with the explicit inclusion of a radial basis and the SO(4) bispectrum obtained through mapping the radial component onto a polar angle of a four dimensional hypersphere.  
With these descriptors, various interatomic potentials for binary Ni-Mo alloys are obtained based on linear and neural network regression models.  
Numerical experiments suggest that both descriptors produce similar results in terms of accuracy. For linear regression, the smooth SO(3) power spectrum is superior to the SO(4) bispectrum when a large band limit is used. In neural network regression, a better accuracy can be achieved with even less number of expansion components for both descriptors.
As such, we demonstrate that spectral neural network potentials are feasible choices for large scale atomistic simulation.
\end{abstract}

\maketitle


\section{Introduction}
The development of accurate and efficient interatomic potentials is a central issue critical in many areas of modern chemical physics. Although \textit{ab-initio} methods such as Kohn-Sham density functional theory (DFT) \cite{Kohn} are accurate and transferable, they are also costly, and therefore limited to applications to systems consisting of only a few thousand atoms.  
On the other hand, empirical force fields are able to handle systems of a much larger scale although accuracy is generally problematic. 
As a consequence, there has been a substantial effort in the last decade to develop efficient and accurate interatomic potentials using machine learning\cite{Zuo2020}.

The development of machine learning interatomic potentials (MLIAPs) has been primarily focused on \textit{feature engineering}, i.e., a numerical descriptor used to represent the local chemical environment for each atomic structure.  
A representation of a chemical environment should be real-valued, unique, invariant to rotation of the system, translation of the system, and permutation of homonuclear atoms \cite{Bartok2013a, Zuo2020}.  
Several representations satisfying these conditions are widely used in fitting MLIAPs, examples of which are: Smooth Overlap of Atomic Positions (SOAP) \cite{Bartok2013a}, Atom-Centered symmetry functions (ACSF) \cite{Behler2007}, Moment Tensor Potentials (MTP) \cite{Shapeev2016}, and Spectral Neighbor Analysis Potential (SNAP) \cite{Thompson2015}.  
Potentials are constructed from these representations through machine learning on \textit{ab-initio} data using regression methods such as generalized linear regression, artificial neural networks, and Gaussian process regression.  
Notable potentials include: the SNAP method which is constructed from the SO(4) bispectrum components and fit using either a linear or quadratic regression \cite{Thompson2015, Wood2018}, Gaussian Approximation Potentials (GAP) constructed using SOAP with Gaussian process regression \cite{bartok2015g}, and High-Dimensional Neural-Networks (NN) potentials constructed using atom-centered symmetry functions with an artificial neural networks \cite{behler2015constructing}. 
For a comprehensive review on descriptor construction and machine learning, please refer to recent literature \cite{Drautz-2019, Willatt-JCP-2019, Behler-JCP-2016, Ceriotti2018}.

Recently, we demonstrated that neural network potentials (NNP) constructed using the SO(4) bispectrum components as the descriptor can achieve good transferability on a rather diverse set of atomic configurations obtained from randomly generated crystalline silicon structures \cite{yanxon2020transferability}.  
In this work, we aim to extend the capability of the NNPs based on the SO(4) bispectrum and smooth SO(3) power spectrum to multicomponent systems as well as provide a comprehensive study of the performance of the SO(4) bispectrum components and the smooth SO(3) power spectrum components as descriptors\cite{Kondor2007, Bartok2013a, Thompson2015}.  
First, we will review the some particular representations of chemical environments related to the harmonic analysis of the atomic neighbor density function, with emphasis on the SO(3) power spectrum and SO(4) bispectrum components. 
In particular, we introduce a different numerical method to compute these descriptors. 
This is followed by a brief discussion on the regression methods used in this study. 
Finally, we apply our approach to a binary-component system Ni-Mo. 
The code that is used in this study is available on \url{https://github.com/qzhu2017/PyXtal_FF}.

\section{Chemical Environment Representations}
A representation of a chemical environment can be considered as a quantitative measure of atomic correlation, or rather, an order parameter, being invariant to translations and rotations of the system as well as permutations of homonuclear atoms.  
First, notice that the spatial distribution of atoms in a chemical environment, up to a cutoff radius ($r_{\textrm{cut}}$), can be represented by a sum of $\delta$ functions.

\begin{equation}\label{density}
    \rho(\bm{r}) = \sum_i^{r_i \leq r_{\textrm{cut}}} \delta(\bm{r}-\bm{r_i})
\end{equation}
\noindent
This is referred to as the atomic neighbor density function\cite{Bartok2013a}.  
The distribution of atoms described by the atomic neighbor density function is not particularly useful by itself, a more useful description of the chemical environment is the angular distribution of atoms in the environment obtained through expanding $\rho(\bm{r})$ as a series on the 2-sphere using spherical harmonics.

\begin{equation*}
    \rho(\bm{r}) = \sum_{l=0}^{+\infty}\sum_{m=-l}^{+l}c_{lm}Y_{lm}(\bm{\hat{r}}),
\end{equation*}
\noindent
where the expansion coefficients $c_{lm}$ are given by:

\begin{equation}\label{expansioncoefs} 
    c_{lm} = \left<Y_{lm}(\bm{\hat{r}})|\rho(\bm{r})\right> = \sum_i^{r_i \leq r_{\textrm{cut}}}Y_{lm}(\bm{\hat{r}_i}).
\end{equation}

For simplicity, we use $\sum_{lm}$ to denote the double summation over $l$ and $m$ from now on. 
Several representations have been constructed using these expansion coefficients.  
Steinhardt constructed his bond order parameters using second and third order combinations of the expansion coefficients (Eq. \ref{expansioncoefs}) to quantify order in liquids and glasses \cite{Steinhardt1983}.
More generally, Kondor constructed an SO(3)-invariant kernel on the 2-sphere using the expansion coefficients of a function defined on the 2-sphere; this kernel provides a method of calculating both the power spectrum and the bispectrum of a function on the 2-sphere \cite{Kondor2007}.  
The SO(3)-invariant power spectrum of Eq. \ref{density} is constructed through taking the autocorrelation of the sequence of expansion coefficients in Eq. \ref{expansioncoefs},

\begin{equation}\label{PS}
    p_l = \sum_{m=-l}^{+l}c_{lm}c^*_{lm}.
\end{equation}

Though $p_l$ from Eq. \ref{PS} satisfies the necessary conditions to represent a chemical environment, it does not carry sufficient information to be useful 
due to the fact that the expansion coefficients in Eq. \ref{expansioncoefs} would not carry any radial information. 
For better application to MLIAPs, Bartok introduced two modifications \cite{Bartok2013a} as follows.

\subsection{Smooth SO(3) Power Spectrum with Explicit Radial Component}

The first modification is to add radial information by expanding $\rho$ not only as a series on the 2-sphere but on a radial basis simultaneously.  
In the second modification, to ensure a smooth similarity kernel, Bartok \cite{Bartok2013a} also expanded Eq. \ref{density} using Gaussians.

\begin{equation}\label{m_density}
    \rho'(\bm{r}) = \sum_i^{r_i \leq r_{\textrm{cut}}} \exp(-\alpha|\bm{r}-\bm{r_i}|^2),
\end{equation}

Then, expanding Eq. \ref{m_density} on the 2-sphere yields
\begin{equation*}
\begin{split}
     \rho'(\bm{r}) &= \sum_{r_i \leq r_{\textrm{cut}}} e^{-\alpha(r^2+r_i^2)} e^{2\alpha\bm{r}\cdot \bm{r_i}}\\
     & =\sum_{r_i \leq r_{\textrm{cut}}} \sum_{lm}  4\pi e^{-\alpha(r^2+r_i^2)} I_l(2\alpha r r_i) Y^*_{lm}(\bm{\hat{r}_i})Y_{lm}(\bm{\hat{r}}),
\end{split}
\end{equation*}
where, $I_l$ is a modified spherical Bessel function of the first kind. The second equation is derived through a spherical harmonic transform of $e^{2\alpha\bm{r}\cdot \bm{r_i}}$.

Radial information can be explicitly added to the representation.  
A convenient radial basis for this purpose proposed by Bartok consists of cubic and higher order polynomials \cite{Bartok2013a}, $g_n(r)$, orthonormalized on the interval $(0, r_{\textrm{cut}})$, while also vanishing at $r_{\textrm{cut}}$,

\begin{equation*}
    \phi_k (r) = (r_{\textrm{cut}} - r)^{k +2}/N_k
\end{equation*}
where
\begin{equation*}
    \begin{split}
        N_k &= \sqrt{\int_0^{r_{\textrm{cut}}} r^2(r_{\textrm{cut}}-r)^{2(k+2)}dr} \\&= \sqrt{\frac{2r_{\textrm{cut}}^{(2k+7)}}{(2k+5)(2k+6)(2k+7)}}
    \end{split}
\end{equation*}

Then orthonormalizing linear combinations of $\phi_k$ from $\phi_1$ up to $\phi_{\textrm{nmax}}$.

\begin{equation}\label{basis}
    g_n(r) = \sum_{k=1}^{n_{\textrm{max}}}W_{nk}\phi_k(r)
\end{equation}

\noindent
$\bm{W}$ is constructed from the overlap matrix $\bm{S}$ by the relation $\bm{W}=\bm{S^{-1/2}}$.  The overlap matrix is given by the inner product \cite{Bartok2013a}: 

\begin{equation}
    \begin{split}
       & S_{pq} = \int_0^{r_{\textrm{cut}}}r^2\phi_p(r)\phi_q(r)dr \\ &= 
        \frac{\sqrt{(2p+5)(2p+6)(2p+7)(2q+5)(2q+6)(2q+7)}}{(5+p+q)(6+p+q)(7+p+q)}
    \end{split}
\end{equation}

In their original work \cite{Bartok2013a}, Bartok \textit{et. al} omitted the $r^2$ term in the integrand of $N_k$ and $S_{pq}$. We included this term to explicitly orthonormalize the radial basis in the spherical polar coordinate system.

Then expanding $\rho'(\bm{r})$ on the 2-sphere and radial basis $g(r)$ in Eq. \ref{basis}, the new expansion coefficients are given by\cite{Bartok2013a}:

\begin{equation}\label{m_expansion}
    \begin{split}
            c_{nlm} &= \left<g_n(r)Y_{lm}(\bm{\hat{r}})|\rho'(\bm{r})\right> \\ 
            &= 4\pi\sum_i^{r_i \leq r_{\textrm{cut}}} e^{-\alpha r_i^2}Y^*_{lm}(\bm{\hat{r}_i})\int_0^{r_{\textrm{cut}}}r^2g_n(r)e^{-\alpha r^2}I_l(2\alpha r r_i)dr
    \end{split}
\end{equation}

The power spectrum components then follow similarly to Eq. \ref{PS}.

\begin{equation}
    p_{n_1 n_2 l} = \sum_{m=-l}^{+l}c_{n_1 l m} c^*_{n_2 l m}
\end{equation}

Note that Bartok further constructed the SOAP kernel to measure the similarity between two chemical environments for Gaussian Process Regression \cite{Bartok2013a}.  
For our purpose, we do not utilize the SOAP kernel itself, but use the smooth SO(3) power spectrum as a descriptor for MLIAPs.

\subsection{SO(4) Bispectrum Components}
An alternative approach to include radial information is to map the atomic neighbor density function within a cutoff radius $r_{\textrm{cut}}$ onto the surface of the four dimensional hypersphere (3-sphere) with a radius of $r_0$ based on the following relations\cite{Bartok2013a, Thompson2015},

\begin{equation*}
    \begin{split}
        s_1 &= r_0\cos\omega \\
        s_2 &= r_0\sin\omega\cos\theta \\
        s_3 &= r_0\sin\omega\sin\theta\cos\phi \\
        s_4 &= r_0\sin\omega\sin\theta\sin\phi,
    \end{split}
\end{equation*}

where the polar angles are defined by:
\begin{equation}\label{polar}
    \begin{split}
        \theta &= \arccos\left(\frac{z}{r}\right)\\
        \phi &= \arctan\left(\frac{y}{x}\right)\\
        \omega &= \frac{\pi r}{r_0}
    \end{split}
\end{equation}

Then, to ensure that the contribution from atoms at $r=r_{\textrm{cut}}$ smoothly goes to zero, it is necessary to augment the atomic neighbor density function (Eq. \ref{density}) with a cutoff function\cite{Thompson2015}, while also including the center atom to avoid unphysical invariance with respect to $\omega$\cite{Bartok2013a}.

\begin{equation}
    \rho(\bm{r}) = \delta(\bm{r}) + \sum_i{f_{\textrm{cut}}(r)\delta(\bm{r}-\bm{r_i})}
\end{equation}

where the cutoff function is defined as\cite{Thompson2015}:

\begin{equation}
    f_{\textrm{cut}}(r) = \begin{cases}
    \frac{1}{2}\left[\cos\left(\frac{\pi r }{r_{\textrm{cut}}}\right) + 1\right],&  r \leq r_{\textrm{cut}}\\
    0,              & r > r_{\textrm{cut}}
    \end{cases}
    \end{equation}

To ensure the mapping produces a one-to-one function defined on the 3-sphere, $r_0$ has to be no smaller than $r_{\textrm{cut}}$.  
For convenience, we simply choose $r_0 = r_{\textrm{cut}}$ to map the atomic neighbor density function onto the entire 3-sphere. 

Now, the atomic neighbor density function mapped onto the 3-sphere by Eq. \ref{polar} can be represented in an expansion of Wigner-$D$ matrix elements in the angle-axis representation, where $2\omega$ is the rotation angle and $\theta,\phi$ define the axis.
\begin{equation*}
    \rho(\bm{r}) = \sum_{j=0}^{+\infty}\sum_{m',m = -j}^{+j}{c^j_{m',m}D^j_{m',m}\left(2\omega;\theta,\phi\right)}
\end{equation*}

\noindent
The Wigner-$D$ matrix elements are mutually orthogonal over the double volume of SO(3) and conveniently the area measure of the 3-sphere corresponds to exactly that (in the angle-axis representation)\cite{VMK}.   
Therefore, the expansion coefficients are obtained by the inner product\cite{Bartok2013a}: 
\begin{equation}
    \begin{split}
    c^j_{m',m} &= \left<D^j_{m',m}|\rho\right> \\&= \int_0^\pi{d\omega \sin^2\omega}\int_0^\pi{d\theta\sin\theta}\int_0^{2\pi}d\phi D^{*j}_{m',m}\left(2\omega;\theta,\phi\right)\rho(\bm{r}) \\&=
    D^{*j}_{m',m}(\bm{0}) + \sum_i f_{\textrm{cut}}(r_i)D^{*j}_{m',m}(\bm{r_i})
    \end{split}
\end{equation}

To obtain the bispectrum components, the triple-correlation of the expansion coefficients is used \cite{Kondor2007}.
The result of which is shown as follows.
\begin{equation}
    \begin{split}
    B_{j_1,j_2,j} &= \sum_{m',m = -j}^{+j}c^{*j}_{m',m}\sum_{m_1',m_1 = -j_1}^{+j_1}c^{j_1}_{m_1',m_1}\times \\ & \sum_{m_2',m_2 = -j_2}^{+j_2}c^{j_2}_{m_2',m_2}C^{jj_1j_2}_{mm_1m_2}C^{jj_1j_2}_{m'm_1'm_2'},
    \end{split}
\end{equation}

where $C$ is a Clebsch-Gordan\cite{VMK} coefficient.

\section{Numerical Implementation}
To fit a MLIAP, both the representation and its gradient are needed.  
The SO(4) bispectrum components, the Smooth SO(3) power spectrum components, and their gradients are implemented in our in house software package PyXtal-FF.  
Although most of the calculations are straightforward as described in the previous section, we will discuss the necessary details where the calculations are nontrivial. 

\subsection{Bispectrum}
For the SO(4) bispectrum components, we need to calculate the Wigner-$D$ matrices for each neighbor.  
Here we use a polynomial form of the Wigner-$D$ matrix elements suggested by Boyle \cite{Boyle}. 

\begin{widetext}
\begin{equation}
    \begin{cases}
    D^j_{m',m} = (-1)^{(j+m)}R_b^{2m}\delta_{-m',m}, & |R_a| < 10^{-15}\\
    D^j_{m',m} = R_a^{2m}\delta_{m',m}, & |R_b| < 10^{-15} \\
    D^j_{m',m} = \sqrt{\frac{(j+m)!(j-m)!}{(j+m')!(j-m')!}}|R_a|^{2j-2m}R_a^{m'+m}R_b^{-m'+m} \times  \sum_{k}\binom{j+m'}{k}\binom{j-m'}{j-m-k}\left(-\frac{|R_b^2|}{|R_a^2|}\right)^k, & |R_a| \geq |R_b|\\
    D^j_{m',m} = (-1)^{j-m} \sqrt{\frac{(j+m)!(j-m)!}{(j+m')!(j-m')!}}R_a^{m'+m}R_b^{m-m'} |R_b|^{2j-2m} \times \sum_{k}\binom{j+m'}{j-m-k}\binom{j-m'}{k}\left(-\frac{|R_a^2|}{|R_b^2|}\right)^k, & |R_a| < |R_b|
    \end{cases}
\end{equation}
\end{widetext}

where $R_a$ and $R_b$ are the Cayley-Klein parameters representing the rotation.  
In the angle-axis representation of rotation the Cayley-Klein parameters representing a rotation about an axis defined by $\bm{r} = (x,y,z)$ through an angle $\omega$ can be written as:

\begin{equation}\label{CK}
    \begin{split}
    R_a &= \cos(\omega/2) + i\frac{\sin(\omega/2)}{r}z \\ 
    R_b &= \frac{\sin(\omega/2)}{r}\left(y+ix\right)
    \end{split}
\end{equation}

These polynomials are finite and the coefficients of each term are known.  
Different from previous works \cite{Bartok2013a, Thompson2015} based on a recursive scheme as discussed in Appendix \ref{app1}, we evaluate the Wigner-$D$ matrix elements using Horner's method for the terms in the summation, which allows evaluation of a polynomial of degree $n$ with only $n$ multiplications and $n$ additions.

\begin{equation}\label{horner}
\begin{split}
        P(x) &= a_0 + a_1x + a_2x^2 + \cdots + a_nx^n \\
             &= a_0 + x(a_1 + x(a_2 + \cdots + x(a_{n-1} + xa_n)))
\end{split}
\end{equation}

Using Horner's method is also convenient for the simultaneous computation of the gradient.  
To obtain the gradient with respect to cartesian coordinates, the chain rule is applied through the Cayley-Klein parameters and their conjugates.

In addition, we make use of the symmetries of the SO(4) bispectrum components discovered by Thompson \cite{Thompson2015}. 
\begin{equation}\label{sym}
    \frac{B_{j_1j_2j}}{2j+1} = \frac{B_{jj_2j_1}}{2j_1+1} = \frac{B_{j_1jj_2}}{2j_2+1}
\end{equation}
\noindent
These symmetries reduce the number of necessary bispectrum components to compute to only the unique components which also greatly reduces the complexity of the gradient calculation.  
For brevity we denote the two inner sums of the bispectrum component calculation as $Z^{m,m'}_{j_1,j_2,j}$\cite{Thompson2015}:

\begin{equation}
     \sum_{m_1, m_1'=-j_1}^{j_1} \sum_{m_2, m_2'=-j_2}^{j_2}c^{j_1}_{m_1',m_1} c^{j_2}_{m_2',m_2} C^{jj_1j_2}_{mm_1m_2}C^{jj_1j_2}_{m'm_1'm_2'}.
\end{equation}

So that, when utilizing the symmetries in Eq. \ref{sym}, the gradient of the bispectrum components with respect to an atom $i$ can be written as\cite{Thompson2015}:
\begin{equation}
    \begin{split}
    \nabla_i B^{(i)}_{j_1,j_2,_j} &= \sum_{m, m'=-j}^{j}\nabla_i\left(c^{j}_{m',m}\right)^* Z^{m,m'}_{j_1,j_2,j} + \\&\frac{2j+1}{2j_1+1}\sum_{m_1, m_1'=-j_1}^{j_1}\nabla_i\left(c^{j_1}_{m_1',m_1}\right)^* Z^{m_1,m_1'}_{j,j_2,j_1} + \\ &\frac{2j+1}{2j_2+1}\sum_{m_2, m_2'=-j_2}^{j_2}\nabla_i\left(c^{j_2}_{m_2',m_2}\right)^* Z^{m_2,m_2'}_{j_1,j,j_2},
    \end{split}
\end{equation}


where the gradient of the inner product with respect to one atom is:
\begin{equation}
    \nabla_ic^j_{m',m} = \nabla_i\left(f_{\textrm{cut}}(r_i)D^{*j}_{m',m}(\bm{r_i})\right)
\end{equation}

\subsection{Smooth SO(3) Power Spectrum}
In calculating the smooth SO(3) power spectrum, three main challenges exist.  
First, the calculation of the spherical harmonics, and second the radial inner product in Eq. \ref{m_expansion}, and third the gradient of the expansion coefficients in Eq. \ref{m_expansion}.  
To start, the spherical harmonics can be considered as a subset of the Wigner-$D$ matrices in the $z$-$y$-$z$ Euler-angle representation, where the spherical harmonic vector $\bm{Y_l}$ is a row vector of the corresponding $D$-matrix $\bm{D^l}$ with some additional scalar factors as given in the equation below\cite{VMK}:

\begin{equation*}
    Y_{lm}\left(\theta,\phi\right) = (-1)^{m}\sqrt{\frac{2l+1}{4\pi}}D^l_{0,-m}\left(\chi, \theta, \phi\right),
\end{equation*}

\noindent
where the Wigner-$D$ matrices are in the $z$-$y$-$z$ Euler-angle representation and $\chi$ is arbitrary, thus, without loss of generality, we choose $\chi=0$. 

The $z$-$y$-$z$ Euler-angle representation represents a rotation about the original $z$-axis through an angle $\alpha$, then a rotation about the new $y$-axis through an angle $\beta$, and then a rotation about the new $z$-axis through an angle $\gamma$, which we can parameterize through composing rotations using the Cayley-Klein parameters in the angle-axis representation.  
For the case of calculating spherical harmonics and choosing $\chi = 0$, we have a rotation about the original y-axis through an angle $\theta$ then a rotation about the new z-axis through an angle $\phi$.  
We represent each of these rotations individually by the Cayley-Klein parameters in the angle-axis representation.

\begin{equation*}
    \begin{split}
        R_{a\theta} &= \cos\frac{\theta}{2}\\
        R_{b\theta} &= \sin\frac{\theta}{2}
    \end{split}
\end{equation*}

\begin{equation*}
    \begin{split}
        R_{a\phi} &= \cos\frac{\phi}{2}+ i\sin\frac{\phi}{2}\\
        R_{b\phi} &= 0
    \end{split}
\end{equation*}

\noindent
To make sense of how to compose rotations represented by the Cayley-Klein parameters it is worthwhile to note that the Cayley-Klein parameters are the matrix elements of the SU(2) representation of rotation.  
So that the rotation (denoted by $\bm{\hat{R}}$) can be represented as:

\begin{equation*}
    \bm{\hat{R}} = \begin{pmatrix}
    R_a & R_b \\
    -R_b^* & R_a^*
    \end{pmatrix}
\end{equation*}

\noindent
Then when composing rotations
\begin{equation*}
    \bm{\hat{R}} = \bm{\hat{R}}_2\bm{\hat{R}}_1
\end{equation*}
\noindent
where $\bm{\hat{R}}_1, \bm{\hat{R}}_2$ are SU(2) matrices that represent arbitrary rotations.  
Performing the matrix multiplication we obtain the composition rule for rotations represented by the Cayley-Klein parameters.

\begin{equation}
    \begin{split}
        R_a &= R_{a2}R_{a1} - R_{b2}R_{b1}^* \\
        R_b &= R_{a2}R_{b1} + R_{b2}R_{a1}^*
    \end{split}
\end{equation}

\noindent
Then for the case of spherical harmonics the composition rule reduces to:

\begin{equation}
    \begin{split}
        R_a &= R_{a\phi}R_{a\theta} = \left(\cos\frac{\phi}{2}+ i\sin\frac{\phi}{2}\right)\cos\frac{\theta}{2}\\
        R_b &= R_{a\phi}R_{b\theta} = \left(\cos\frac{\phi}{2}+ i\sin\frac{\phi}{2}\right)\sin\frac{\theta}{2}
    \end{split}
\end{equation}

\noindent
Finally, using the composition rule we can then calculate the spherical harmonics using their relationship to the Wigner-$D$ matrices.  
\begin{equation}
    Y_{lm}\left(R_a, R_b\right) = (-1)^m\sqrt{\frac{2l+1}{4\pi}}D^l_{0,-m}(R_a, R_b)
\end{equation}

The radial inner product $\int_0^{r_{\textrm{cut}}}r^2g_n(r)e^{-\alpha r^2}I_l(2\alpha r r_i)dr$ in Eq. \ref{m_expansion} cannot be solved analytically so we employ numerical integration for this purpose.  
Chebyshev-Gauss quadrature is used so that the quadrature nodes for the interval $(0, r_{\textrm{cut}})$ never include $r=0$ for any $N$ number of nodes in the quadrature; the Chebyshev-Gauss quadrature nodes for the interval $(0, r_{\textrm{cut}})$ are given by:

\begin{equation}
    x_i = \frac{r_{\textrm{cut}}}{2}\left[\cos\left(\frac{2i-1}{2N}\pi\right)+1\right]
\end{equation}

Avoiding the removable singularity at $r=0$ due to $I$ allows for the use of the following recursion relation to compute $I$ at each of the nodes.  

\begin{equation}
\begin{cases}
    I_0(x) &= \frac{\sinh(x)}{x} \\
    I_1(x) &= \frac{x\cosh(x)-\sinh(x)}{x^2}\\
    ~~~\vdots \\
    I_n(x) &= I_{n-2}(x)-\frac{2n-1}{x}I_{n-1}(x)
\end{cases}
\end{equation}

The gradient of the smooth SO(3) power spectrum components then follows:

\begin{equation}
     \nabla_ip_{nn'l} =  \sum_{m=-l}^{+l}\left(c^*_{n'lm}\nabla_ic_{nlm}+c_{nlm}\nabla_ic^*_{n'lm}\right)
\end{equation}

where the gradient of the expansion coefficients is obtained through the applying the product rule on Eq. \ref{m_expansion} and then differentiating under the integral sign (as $r_i$ is independent of $r$).

\begin{equation}
    \begin{split}
        \nabla_i & c_{nlm} =\\ & 4\pi\nabla_i\left(e^{-\alpha r_i^2}\right)Y^*_{lm}(\bm{\hat{r}_i})\int_0^{r_{\textrm{cut}}}r^2g_n(r)e^{-\alpha r^2}I_l(2\alpha r r_i)dr + \\
        & 4\pi e^{-\alpha r_i^2}\nabla_i\left(Y^*_{lm}(\bm{\hat{r}_i})\right)\int_0^{r_{\textrm{cut}}}r^2g_n(r)e^{-\alpha r^2}I_l(2\alpha r r_i)dr + \\
        & 4\pi e^{-\alpha r_i^2}Y^*_{lm}(\bm{\hat{r}_i})\nabla_i\left(\int_0^{r_{\textrm{cut}}}r^2g_n(r)e^{-\alpha r^2}I_l(2\alpha r r_i)dr\right)
    \end{split}
\end{equation}
\begin{equation*}
    \nabla_i\left(e^{-\alpha r_i^2}\right) = -2\alpha r_i e^{-\alpha r_i^2} \bm{\hat{r}_i}
\end{equation*}

\begin{equation*}
    \begin{split}
            \nabla_i\left(\int_0^{r_{\textrm{cut}}}r^2g_n(r)e^{-\alpha r^2}I_l(2\alpha r r_i)dr\right) =\\
            2\alpha\int_0^{r_{\textrm{cut}}}r^3g_n(r)e^{-\alpha r^2}I_l'(2\alpha r r_i)dr \bm{\hat{r}_i}
    \end{split}
\end{equation*}

\noindent
We again evaluate the radial integral using Chebyshev-Gauss quadrature and use the following recursion relation for the evaluation of the first derivative of the modified spherical Bessel function.

\begin{equation*}
    I_n'(x) = \frac{1}{2n+1}[nI_{n-1}(x)+(n+1)I_{n+1}(x)].
\end{equation*}

Computing the gradient of the spherical harmonics is not as trivial as computing the gradient of the Wigner-$D$ functions due to the singularities that exist at the north and south poles of the 2-sphere in Cartesian and spherical polar coordinates.  
Here we remove those singularities through taking the gradient with respect to the covariant spherical coordinates.  
The covariant spherical coordinates are related to Cartesian coordinates by the following relation\cite{VMK}:

\begin{equation*}
    \begin{split}
        x_{+1} &= -\frac{1}{\sqrt{2}}\left(x+iy\right) \\
        x_{0} &= z \\
        x_{-1} &= \frac{1}{\sqrt{2}}\left(x-iy\right)
    \end{split}
\end{equation*}

\noindent
Then, the gradient of the spherical harmonics with respect to the covaraint spherical coordinates is given by\cite{VMK}:

\begin{equation}
    \begin{split}
        \nabla_0Y_{l,m} &= -\frac{l}{r}\sqrt{\frac{(l+1)^2-m^2}{(2l+1)(2l+3)}}\times Y_{l+1,m} \\
         &- \frac{l+1}{r}\sqrt{\frac{l^2-m^2}{(2l-1)(2l+1)}}\times Y_{l-1,m} \\ \\
         \nabla_{\pm 1} Y_{l,m} &= -\frac{l}{r}\sqrt{\frac{(l\pm m + 1)(l\pm m +2)}{2(2l+1)(2l+3)}} \times Y_{l+1,m\pm 1} \\
         &- \frac{l+1}{r}\sqrt{\frac{(l \mp m -1)(l\mp m)}{2(2l-1)(2l+1)}}\times Y_{l-1,m\pm 1}
    \end{split}
\end{equation}

\noindent
So that we can obtain the gradient with respect to Cartesian coordinates by transforming the basis vectors back to Cartesian unit vectors\cite{VMK}.

\begin{equation}
    \begin{split}
        \bm{e_x} &= \frac{1}{\sqrt{2}}\left(\bm{e_{-1}}-\bm{e_{+1}}\right) \\
        \bm{e_y} &= \frac{i}{\sqrt{2}}\left(\bm{e_{-1}}+\bm{e_{+1}}\right) \\
        \bm{e_z} &= \bm{e_0}
    \end{split}
\end{equation}

\section{Interatomic Potential Fitting}
In the present work, we adopted two fitting approaches: neural networks and linear regressions. 
Both techniques predict collection of atomic energies of a given structure: $E_\textrm{total} = \Sigma_i E_i$, where $i$ loops through all atoms in the structure. 
Each $E_i = f(\boldsymbol{X}_i)$ is a function of descriptors, $\bm{X}_i$, representing the chemical environment around $\boldsymbol{r}_i$, a set of atomic positions relative to the $i$-th center atom within a cutoff radius ($r_{\textrm{cut}}$).  
Since the atom-centered descriptors are derived analytically as shown in the previous section, one can deduce explicit forms of the functional to calculate forces and stress tensors.

\subsection{Linear Regression}

Given the atom-centered descriptors ($\boldsymbol{X}_i$), the functional form for $E_i$ can be expressed as a linear combination of the descriptors:
\begin{equation}\label{PolyEnergy}
    E_{\textrm{total}} = \sum^{N}_{i=1} E_i = \theta_0 + \boldsymbol{\theta} \cdot \sum^{N}_{i=1}\boldsymbol{X}_i,
\end{equation}
where $\theta_0$ and $\boldsymbol{\theta}$ denote as the weight parameters, and $N$ is the total atoms in the structure.  
The forces on each atom can be obtained by computing the partial derivative of $-\partial E/\partial \bm{r_i}$ through the chain rule. Optionally, one can also include information on virial stress in the training.  
In the context of linear regression, the objective is then to minimize the overall errors with respect to energy, forces and stresses between the linear model and the training samples. To prevent overfitting, a penalty function, usually the $l_1$ or $l_2$ norm of $\boldsymbol{\theta}$  can be added to the expression of loss function to serve as a regularization term. Therefore, the final expression is
\begin{equation}\label{loss}
    \Delta = \Bar{E}_{\textrm{mse}} + \beta \Bar{F}_{\textrm{mse}} + \gamma \Bar{\sigma}_{\textrm{mse}} +
    \lambda ||\boldsymbol{W}||_n,
\end{equation}
where the $\Bar{E}_{\textrm{mse}}, \Bar{F}_{\textrm{mse}}, \Bar{\sigma}_{\textrm{mse}}$ denote the mean squared errors due to energy, force and virial stress, $||\boldsymbol{W}||_n$ denotes the $n$-norm of the weight vector, and $\beta, \gamma, \lambda$ denote the coefficients to balance the emphasis of training on force, stress and penalty function. For the case of linear regression, $\boldsymbol{W}$ is the concatenated vector of $\{\theta_0, \boldsymbol{\theta}\}$. 

\subsection{Neural Network Regression}

In a NN regression, the atom-centered descriptors serve as the inputs to the first layer of the neural network architecture.  
The NN architecture also consists of output layer and hidden layers, where hidden layers reside in between input and output layers.  
Within a layer, there are collection of units or nodes called neurons.  
The connectivity between these neurons in the layers mimics synapses of neurons in a biological structure.  
The signals or atom-centered descriptors permeate into the hidden layer to the output neuron in the following general form:
\begin{equation}\label{neuron}
    X^{l}_{n_i} = a^{l}_{n_i}\bigg( b^{l-1}_{n_i} + \sum^{N}_{n_j=1} W^{l-1, l}_{n_j, n_i} \cdot X^{l-1}_{n_j} \bigg)
\end{equation}
The neuron $X_{n_i}^l$ at $l$-th layer is established by the relationships between the weight parameter $W^{l-1, l}_{n_j, n_i}$, the bias parameter $b^{l-1}_{n_i}$, and the neurons in the prior layers $X^{l-1}_{n_j}$. Here, $W^{l-1, l}_{n_j, n_i}$ specifies the connectedness of the $n_j$ neuron at $(l-1)$-th layer to the neuron $n_i$ at $l$-th layer.  
Then, an activation function $a_{n_i}^l$ is applied to the process for the purpose of introducing non-linearity to the neurons. $X_{n_i}$ at the output layer is equivalent to an atomic energy in the scope of this study, in which the collection of these atomic energies are the total energy of the system.  
The details about NN architecture and its application in interatomic potential fitting have been discussed in many excellent review works recently \cite{behler2015constructing, Behler-JCP-2016,Artrith2016}.

\section{Results and Discussion}
In this section, we will first compare the computational costs for each descriptor calculation as a function of the hyperparameters. 
The accuracy of each representation in relation to both the number of descriptors and its computational cost will be then investigated by regressing on energies, forces, and stresses of a representative binary alloy Ni$_3$Mo/Ni$_4$Mo system using linear regression.  
Last, we will introduce a more flexible NN regression model to improve the accuracy of fitting on the extended Ni-Mo data set within a larger chemical space. 

In parallel to force field fitting, generating a diverse training data set is also a challenging task. Recently, there is an increasing trend for research groups to share their own data to the entire MLIAP community. Thanks to this trend, we choose to examine the data set from a recent work by Li \textit{et al} \cite{Li-PRB-2018}, which includes 4019 atomic configurations for elemental Ni, Mo, Ni$_3$Mo, Ni$_4$Mo, and doped Ni-Mo alloys.  
The training dataset consists of (1) undistorted ground state structures for Ni, Mo, $\textrm{Ni}_3\textrm{Mo}$, and $\textrm{Ni}_4\textrm{Mo}$, (2) distorted structures obtained by applying strains of $-10\%$ to $10\%$ at $1\%$ intervals to a bulk supercell, (3) surface structures of elemental structures, (4) snapshots from ab initio molecular dynamics simulations of the bulk supercell at several temperatures, (5) doped alloy structures constructed by partial substitution of the bulk fcc Ni with Mo and the bulk bcc Mo with Ni. In addition, we also used the extra dataset on Mo from Ref. \cite{Chen2017}. For the computation of each descriptor below, we used a uniform cutoff distance of 4.9 \AA.

\subsection{Computational Cost Comparison}

We begin with evaluating the computational cost of the SO(4) bispectrum components and the Smooth SO(3) power spectrum components, which requires some measure of the cost of each method.  
By far, the gradient is the most expensive part of the calculation so we estimate the cost of each method by the accumulation of the gradient for one neighbor.  
The cost function for each method is evaluated by the asymptotic cost of accumulating the gradient plus the cost of precomputing the expansion coefficients and their gradients for a given truncation.  
For the SO(4) bispectrum components, the cost of precomputation is equivalent to the number of Wigner-$D$ matrix elements to evaluate, $\sum_{j=0}^{2j_\textrm{max}}(j+1)^2$, where for the smooth SO(3) power spectrum the cost of precomputation is equal to the number of Wigner-$D$ matrix elements to evaluate, $(l_{\textrm{max}}+2)^2$, added to the number of radial functions to evaluate for the quadrature (Eq. \ref{costfun}), to compute each integral, we use $10(n+l+1)$ quadrature nodes.  In our implementation, the cost of evaluating the radial functions is less than that of evaluating the $D$-functions although for the sake of simplicity of the cost model we treat these costs as equal.

\begin{equation}
        \textrm{cost} = \textrm{cost}_{\textrm{accum}} + \textrm{cost}_{\textrm{precomputation}}
\end{equation}

Therefore, we estimate the computational cost of each descriptor as follows,
\begin{equation}\label{costfun}
    \begin{split}
        \textrm{SO(4):~} & j_{\textrm{max}}^5 + \sum_{j=0}^{2j_{\textrm{max}}}{(j+1)^2} \\
        \textrm{SO(3):~} & n_{\textrm{max}}^2 l_{\textrm{max}}^2 + \left[(l_{\textrm{max}}+2)^2 + \sum_{n=1}^{\textrm{n}_\textrm{max}}\sum_{l=0}^{\textrm{l}_\textrm{max}}{10(n+l+1)} \right]       
    \end{split}
\end{equation}

The cost of each descriptor is then compared with the number of elements of that descriptor.  
The number of unique elements of each descriptor are given by:

\begin{equation}\label{ndes}
    \begin{split}
        N_{\textrm{SO(4)}} &= (j_{\textrm{max}}+1)(j_{\textrm{max}}+2)(j_{\textrm{max}}+3/2)/3\\
        N_{\textrm{SO(3)}} &= n_{\textrm{max}}(n_{\textrm{max}}+1)(l_{\textrm{max}}+1)/2
    \end{split}
\end{equation}

\begin{figure}[h]
    \centering
    \includegraphics[width=10cm]{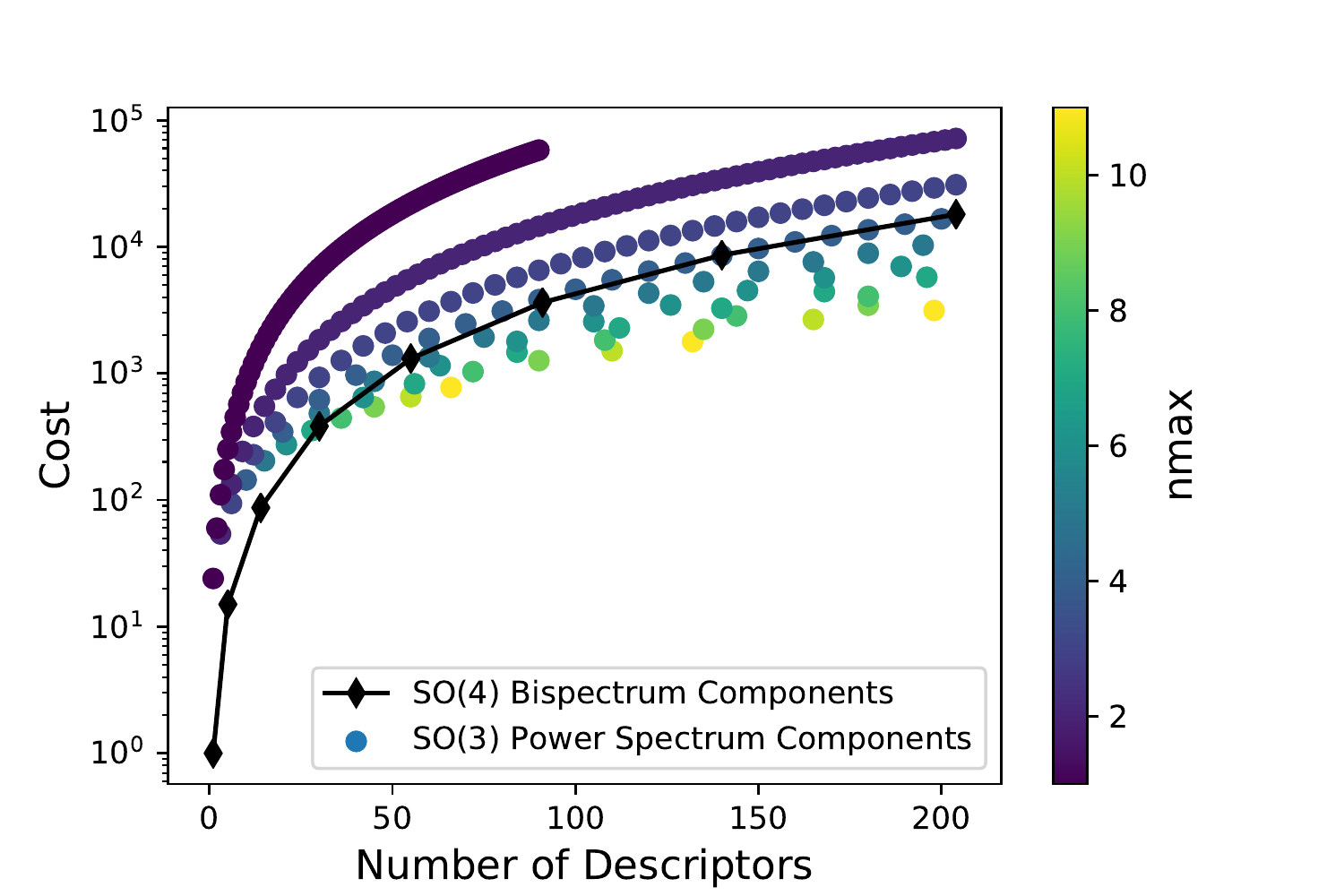}
    \caption{The computational cost of the SO(4) bispectrum descriptor and the smooth SO(3) power spectrum descriptor versus the total number of elements of that descriptor. The smooth SO(3) power spectrum is also colored according to the number of radial components in the expansion.}
    \label{cost}
\end{figure}

In Fig. \ref{cost}, we plot the computational cost given by Eq. \ref{costfun} with respect to the number of descriptors (Eq. \ref{ndes}) for both SO(4) bispectrum and SO(3) power spectrum. Clearly, we find that in the low band limit $(N\leq30)$, the SO(4) bispectrum components are much less costly than the Smooth SO(3) power spectrum components, where at higher band limits, including more terms in the radial expansion of the smooth SO(3) power spectrum results in a less costly computation in comparison to the SO(4) bispectrum components.

\subsection{Linear regressions on $\textrm{Ni}_{4}\textrm{Mo}$ and $\textrm{Ni}_{3}\textrm{Mo}$}

To evaluate the performance of these two descriptors, we first choose a subset of data from the Ni-Mo dataset, which includes 642 atomic configurations only in the Ni$_3$Mo and Ni$_4$Mo stoichiometries. 
We then fit linear regressions to this data for each representation using a set of descriptors obtained through different hyperparameters in Eq. \ref{costfun}, while varying the coefficients of force's contribution to the total loss function.  

\begin{figure}[ht]
    \centering
    \includegraphics[width=0.5\textwidth]{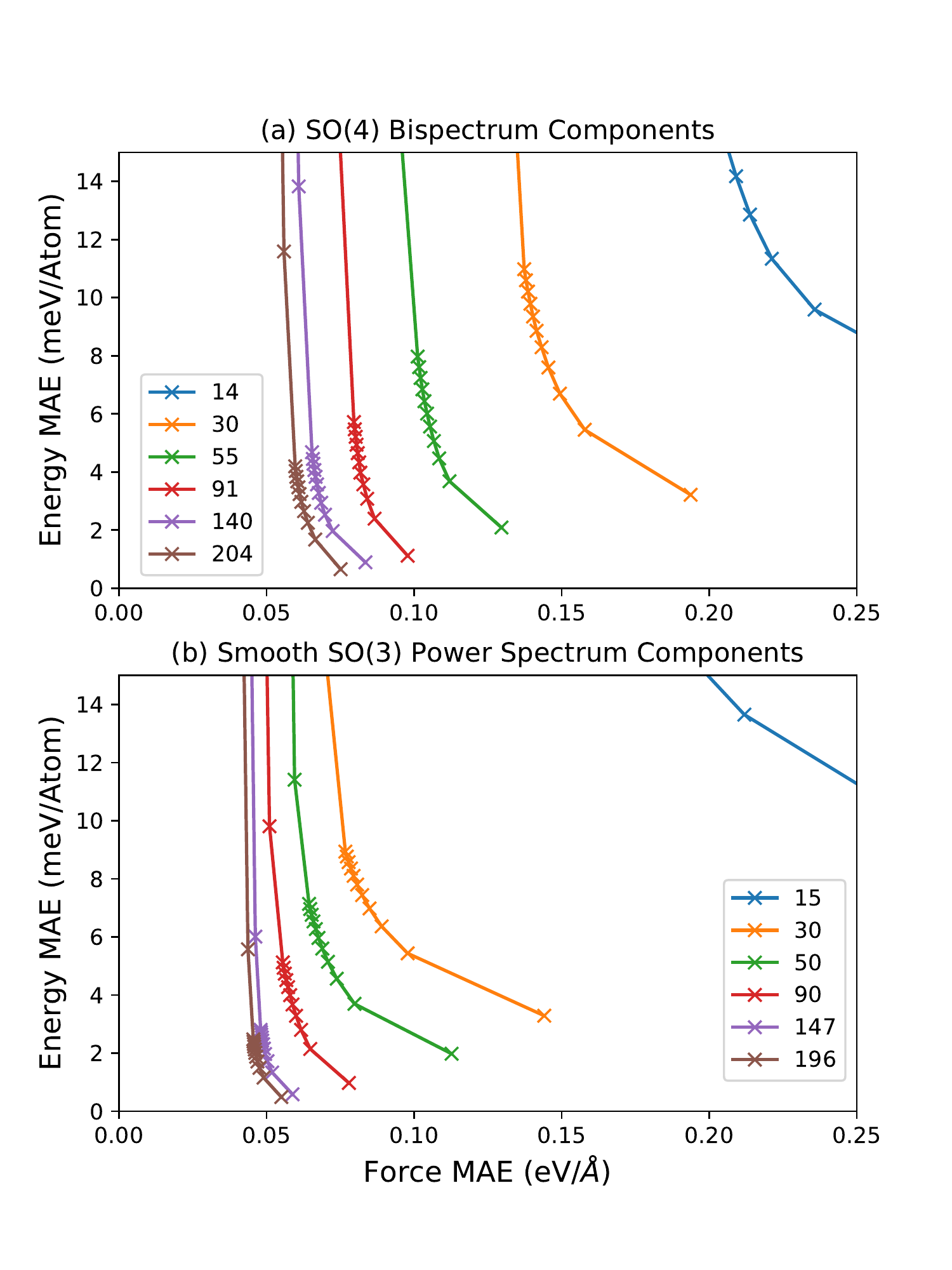}
    \caption{Linear regressions of both the SO(4) and SO(3) representations with varying numbers of components. The force coefficients used fall between 1e-6 and 1e+0 with most points falling between 1e-5 and 1e-4.}
    \label{LR}
\end{figure}

The results of these regressions are shown in Figure \ref{LR}. Clearly, there is a general trend that both SO(3) power spectrum and SO(4) bispectrum can continuously achieve better accuracy with the inclusion of more components, although at high bandlimits that increased accuracy becomes marginal. 
In addition, the results show that high bandlimit fits vary less with respect to the change of force coefficient, indicating a convergence of the regression. However, a full convergence at high bandlimits results an in incredibly expensive calculations.  In real applications, it is generally advised to choose a smaller bandlimit. 
For the SO(4) bispectrum components, holding the truncation of $j_\textrm{max}=3$ is a rather common choice \cite{Thompson2015, Zuo2020, Li-PRB-2018}. 
A more detailed analysis regarding the cost of computing the SO(4) bispectrum components with respect to $j_\textrm{max}$ can be found in Ref. \cite{Wood2018}. 
Therefore, we aim to for a better solution through investigating the smooth SO(3) power spectrum.

Indeed, we find that the smooth SO(3) power spectrum components converge more quickly to lower errors in comparison to the SO(4) bispectrum components while also converging to a lower error overall. 
For instance, using only $90$ smooth SO(3) power spectrum components yields similar accuracy (2.14 meV/atom in energy MAE and 0.06 eV/\AA~ in force MAE) to $204$ bispectrum components (1.68 meV/atom in energy MAE and 0.07 eV/\AA~ in force MAE) if we hold the force coefficient at 1e-5. 
To further illustrate the performance of both descriptors in terms of computational cost, we calculate both the 90 component smooth SO(3) power spectrum and 204 component SO(4) bispectrum for the ground state Ni$_3$Mo structure with 8 atoms in the unit cell at a cutoff radius of $4.9$ \AA~ with the gradient; the smooth SO(3) power spectrum component calculation is completed 0.56 seconds whereas the SO(4) bispectrum component calculation is completed in 3.97 seconds. 
Since our code is written in Python (using the LLVM compiler through Numba \cite{numba}), we expect the run time will be less if the code is rewritten in C++ or Fortran. 
These results suggest that the smooth SO(3) power spectrum is a more efficient descriptor in terms of both accuracy and computational cost in the context of linear regression.

Although both descriptors yield satisfactory accuracy on the Ni$_3$Mo/Ni$_4$Mo data set, we found it hard to maintain the same level of accuracy when extending the training dataset with other stoichimetries (e.g., elemental Ni/Mo) for the regression.  
In principle, one can improve the regression by tuning force and stress coefficients, applying regularization, and adopting a nonuniform weight scheme on each sample\cite{Li-PRB-2018},\cite{Wood2018}. 
However, a more automated approach to dealing with large data is to employ a more flexible regression model such as NN regression to be presented in the following subsection.

\begin{table*}[ht]
\caption{Comparison of the Spectral Neural Networks Models' MAE values from different descriptors. For reference, the previous NiMo model trained from SNAP \cite{Li-PRB-2018} is also included. Note that in the SNAP model \cite{Li-PRB-2018}, only 247 Mo structures were used for training. In our work, we replaced the elastic configuration data with the data set from Ref. \cite{Chen2017}. In the parenthesis it gives the number of configurations for each group.}
\centering
\begin{tabular}{c|cccccccccccc}
\hline
\hline
~~~~Properties~~~~& ~Descriptor~ &~$j_\textrm{max}$~& ~$l_\textrm{max}$~&~$n_\textrm{max}$~& ~~~~Architecture~~~~ & ~~~Mo~~~  & ~~~Ni~~~  & ~~Mo$_\textrm{Ni}~~$ & ~~Ni$_\textrm{Mo}$~~ & ~~Ni$_3$Mo~~ & ~~Ni$_4$Mo~~ & ~~Overall~~ \\ 
             &        && &  &  & (377) & (414) & (918) & (1668) & (321) & (321) & (4019) \\ \hline

\multicolumn{1}{c|}{\multirow{4}{*}{\begin{tabular}[c]{@{}c@{}}Energy\\ (meV/atom)\end{tabular}}} & SO(4)\cite{Li-PRB-2018} &3& & & Linear Reg. &16.2     & 7.9      & 22.7         & 33.9          & 5.2         & 4.0           & 22.5           \\ 
\multicolumn{1}{c|}{} & SO(4) &3&& &30-16-16-1& 6.2 & 7.3 & 5.6 & 6.1 & 6.1 & 6.4 & 6.1 \\ 
\multicolumn{1}{c|}{} & SO(3) &&4& 3&30-16-16-1& 6.3 & 3.6  & 6.2  & 6.7  & 4.9  & 4.6  & 5.9  \\ \hline

\multicolumn{1}{c|}{\multirow{4}{*}{\begin{tabular}[c]{@{}c@{}}Force\\ (eV/\AA)\end{tabular}}} 
& SO(4)\cite{Li-PRB-2018} &3&&& Linear Reg.& 0.29     & 0.11     & 0.13        & 0.55          & 0.16        & 0.14        & 0.23           \\ 
& SO(4) &3&& &30-16-16-1 & 0.19 & 0.07 & 0.06 & 0.10  & 0.10  & 0.09 & 0.10  \\ 
& SO(3) &&4& 3&30-16-16-1 & 0.18  & 0.04 & 0.06 & 0.10 & 0.09  & 0.07 & 0.08 \\ \hline\hline
\end{tabular}\label{t1}
\end{table*}

\begin{table}
\caption{Comparison of elastic properties predicted from several different Models. $B$ and $G$ denote the empirical Voigt-Reuss-Hill average of bulk and shear moduli respectively. $\nu$ is the Poisson's ratio.}
\centering
\begin{tabular}{lcccc}
\hline\hline
\multicolumn{1}{l}{}           & ~~~~~~DFT~~~~~~  & ~~SNAP\cite{Li-PRB-2018}~~  & ~~~~SO(4)~~~~ & ~~SO(3)~~ \\ \hline
\multicolumn{1}{l|}{$\sigma$(MAE) (GPa)}& & N/A    & 0.295       & 0.289    \\ \hline
\multicolumn{5}{l}{Mo}                                                            \\ \hline
\multicolumn{1}{l|}{$c_{11}$ (GPa)}  & 472  & 475    & 487      & 479      \\
\multicolumn{1}{l|}{$c_{12}$ (GPa)}  & 158  & 163    & 153      & 168     \\
\multicolumn{1}{l|}{$c_{44}$ (GPa)}  & 106  & 111    & 108      & 82    \\
\multicolumn{1}{l|}{$B$ (GPa)}       & 263  & 267    & 265      & 271     \\
\multicolumn{1}{l|}{$G$ (GPa)}       & 124  & 127    & 129      & 106     \\
\multicolumn{1}{l|}{$\nu$}           & 0.30 & 0.29   & 0.29     & 0.33     \\ \hline
\multicolumn{5}{l}{Ni}                                                            \\ \hline
\multicolumn{1}{l|}{$c_{11}$ (GPa)}  & 276  & 269    & 275      & 271      \\
\multicolumn{1}{l|}{$c_{12}$ (GPa)}  & 159  & 150    & 162      & 150     \\
\multicolumn{1}{l|}{$c_{44}$ (GPa)}  & 132  & 135    & 137      & 120    \\
\multicolumn{1}{l|}{$B$ (GPa)}       & 198  & 190    & 199      & 188    \\
\multicolumn{1}{l|}{$G$ (GPa)}       & 95   & 97     & 96       &  88   \\
\multicolumn{1}{l|}{$\nu$}           & 0.29 & 0.28   & 0.29     & 0.30     \\ \hline
\multicolumn{5}{l}{Ni$_3$Mo}                                                         \\ \hline
\multicolumn{1}{l|}{$c_{11}$ (GPa)}  & 385  & 420    & 426      & 402      \\
\multicolumn{1}{l|}{$c_{22}$ (GPa)}  & 402  & 360    & 354      & 382    \\
\multicolumn{1}{l|}{$c_{33}$ (GPa)}  & 402  & 408    & 379      & 394    \\
\multicolumn{1}{l|}{$c_{12}$ (GPa)}  & 166  & 197    & 159      & 159    \\
\multicolumn{1}{l|}{$c_{13}$ (GPa)}  & 145  & 162    & 133      & 109    \\
\multicolumn{1}{l|}{$c_{23}$ (GPa)}  & 131  & 145    & 208      & 173     \\
\multicolumn{1}{l|}{$c_{44}$ (GPa)}  & 58   & N/A    & 54       & 70      \\
\multicolumn{1}{l|}{$c_{55}$ (GPa)}  & 66   & N/A    & 68       & 52      \\
\multicolumn{1}{l|}{$c_{66}$ (GPa)}  & 94   & 84     & 79       & 58    \\
\multicolumn{1}{l|}{$B$ (GPa)}       & 230  & 243    & 240      & 229      \\
\multicolumn{1}{l|}{$G$ (GPa)}       & 89   & 100    & 80       & 80      \\
\multicolumn{1}{l|}{$\nu$}           & 0.33 & 0.32   & 0.35     & 0.34     \\ \hline
\multicolumn{5}{l}{Ni$_4$Mo}                                            \\ \hline
\multicolumn{1}{l|}{$c_{11}$ (GPa)}  & 313  & 326    & 319      & 343      \\
\multicolumn{1}{l|}{$c_{33}$ (GPa)}  & 300  & 283    & 294      & 293     \\
\multicolumn{1}{l|}{$c_{12}$ (GPa)}  & 166  & 179    & 166      & 160    \\
\multicolumn{1}{l|}{$c_{13}$ (GPa)}  & 186  & 164    & 199      & 193    \\
\multicolumn{1}{l|}{$c_{44}$ (GPa)}  & 130  & 126    & 136      & 131    \\
\multicolumn{1}{l|}{$c_{66}$ (GPa)}  & 106  & N/A    & 102      & 113    \\
\multicolumn{1}{l|}{$B$ (GPa)}       & 223  & 220    & 221      & 222    \\
\multicolumn{1}{l|}{$G$ (GPa)}       & 91   & 95     & 96       & 102    \\
\multicolumn{1}{l|}{$\nu$}           & 0.33 & 0.31   & 0.31     & 0.30    \\ \hline\hline
\end{tabular}
\label{t2}
\end{table}

\subsection{Neural network regressions on Ni-Mo alloys}

When dealing with a large amount of data, linear regression requires very fine tuning of hyperparameters to achieve acceptable accuracies.  
To achieve these accuracies, optimization schemes are adopted to adjust hyperparameters such as descriptor size, specie weights, cutoff radii, and nonuniform data weighting so that obtaining an optimal fit requires many training cycles \cite{Li-PRB-2018,Chen2017,Wood2018}.  
NN regression provides a more automated approach to achieve greater accuracy on larger datasets without the need for high bandlimit descriptors or heavy hyperparameter optimization.  
In this study, we seek to use a small set of descriptors (30) to train a MLIAP on the entire Ni-Mo dataset consisting of over 4000 structures to satisfactory accuracy through a simple feed forward neural network consisting of two hidden layers of 16 neurons each. For a fair comparison, we prepare two sets of descriptors: (1) the bispectrum components with $j_\textrm{max}$ = 3; and (2) the smooth SO(3) power spectrum components with $l_\textrm{max}$ = 4 and $n_\textrm{max}$ = 3. To ensure that the results can describe elastic deformation well, we also consider the virial stresses for the elastic configurations in the training. Correspondingly, we set the $\beta$=3e-3, $\gamma$=1e-4, and $\lambda$=1e-8 for the evaluation of the loss functions (Eq. \ref{loss}) in all subsequent NN runs. 

Table \ref{t1} lists the training results in terms of energy and force for all three models. In the previously reported linear SNAP model \cite{Li-PRB-2018}, the overall fitting results are 22.5 meV/atom in energy MAE, and 0.23 eV/{\AA}. Clearly, both NN models are able to yield significantly better results (~6 meV/atom for energy and 0.08 eV/{\AA} for force) than the previous reported linear model. Notably, the linear regression also reports drastically lower accuracy in both energy and force for the $\textrm{Mo}_\textrm{Ni}$/$\textrm{Mo}_\textrm{Ni}$ sets, suggesting that the elemental Ni/Mo and Ni$_3$Mo/Ni$_4$Mo portions of the data were weighted much higher in the regression. In particular, the 1668 Ni$_\textrm{Mo}$ set, occupying the largest percentage of the data, has a energy MAE of 33.9 meV/atoom and force MAE of 0.55 eV/{\AA}. As such, the predictability of linear SNAP model is likely to be limited in describing the configurations in the vicinity of the $\textrm{Mo}_\textrm{Ni}$/$\textrm{Mo}_\textrm{Ni}$ alloys.  
In contrast, the neural network regressions do not need a special weighting scheme. The models from both SO(4) bispectrum and SO(3) power spectrum yield not only lower energy and force errors for the overall fitting. The energy/force errors for each group are also more evenly distributed. 

The elastic tensor is another important metric to check if the trained MLIAPs are able to reproduce the fine details of the PES on the representative basins. To ensure a satisfactory fitting to the elastic properties, we also included training on the stress tensors for the elastic configurations from the previous works \cite{Li-PRB-2018, Chen2017}. Table \ref{t2} shows the predicted elastic properties from each model for the ground state structures of BCC Mo, FCC Ni, Ni$_3$Mo, and Ni$_4$Mo. In agreement with the previously reported linear SNAP model \cite{Li-PRB-2018}, the elastic data predicted by each MLIAP agrees with the reported DFT result within similar levels of accuracy across all four ground state structures. In the previous work, it is likely that the authors adjusted the weight for each group of structures in order to achieve a better fit in the elastic properties at the expense of accuracy in energy and force. However, these NN regressions can circumvent this trade-off by using a more flexible expression in describing the target properties (energy, force, stress tensor) in fitting. As such, the NN models can yield greater accuracy with respect to energy and force while maintaining accuracy in elastic properties all without the need for heavy hyperparameter optimization. 

Last, it is also of interest to compare the performance of fitting between the SO(4) bispectrum and smooth SO(3) power spectrum models. In the previous section, it is clear that SO(3) is superior to SO(4) in the context of linear regression. However, this is no longer the case for NN regression. With the same number of descriptors (30), both NN models yield very similar levels of accuracy. In terms of elastic properties prediction, the SO(4) model seems to be slightly better than SO(3) though SO(3) generated a slightly lower MAE value for stress tensors overall \footnote{We note that each NN training follows a stochastic optimization process. So each time, it may generate slightly different results. Hence the comparison is not definitive.}. From the point view of computational cost, computing the 30 bispectrum components is less expensive than computing the same number of power spectrum components. Therefore, it is fair to conclude that two descriptors are competitive for the application of NN regression.

\section{Conclusion}
In summary, we present a numerical implementation of computing the atom-centered descriptors derived from harmonic analysis, which include the SO(4) bispectrum components and the smooth SO(3) power spectrum.  
Using these descriptors to fit machine learning interatomic potentials for a small set of Ni-Mo stoichiometries within a narrow chemical composition space, we found that both descriptors are able to yield satisfactory accuracy within the framework of linear regression.   
However, the linear regression is not easily extended to fit a more diverse data set from a larger chemical composition space and even then accuracy can still be lacking without hyperparameter optimization such as descriptor size, specie weights, cutoff radii, and nonuniform data weighting.  
Hence, we demonstrate that neural networks regression paired with the SO(4) bispectrum components or the smooth SO(3) power spectrum components can provide a better trained model without the need for large band limit descriptors or heavy hyperparameter optimization.
The validity of the trained models are further supported by the accuracy of elastic property calculations. 
Last, the SO(3) power spectrum descriptor clearly exhibits better agreement with the total energy than the SO(4) bispectrum components, thus it is a better choice for linear regression. However, when adopted to the neural networks regression, both descriptors tend to yield the same level of accuracy. A further comparison on the performances of different types of descriptors will the be subject of future study.

\section*{Acknowledgments}
We acknowledge the NSF (I-DIRSE-IL: 1940272) and NASA (80NSSC19M0152) for financial support. The computing resources are provided by XSEDE (TG-DMR180040). The authors thank Dr. A. Thompson (Sandia), Dr. S. Ong (UCSD) and Dr. Y-G Li (USCD) for insightful discussions in the computation of bispectrum coefficients.

\section*{Data Availability}
The data that support the findings of this study are available from the corresponding author upon reasonable request. The source code to analyze the data is available on \url{https://github.com/qzhu2017/PyXtal_FF}.

\appendix
\section{Alternative expression to compute $D$} \label{app1}
We are aware that two previous works \cite{Bartok2013a, Thompson2015} used a different approach to compute the Wigner-$D$ martices\cite{Bartok2013a,Thompson2015}.
To start, a different set of Cayley-Klein parameters were used,
\begin{equation}
    \begin{split}
        R_a &= \frac{1}{\sqrt{r^2+r^2\cot^2(\omega/2)}}\left(r\cot(\omega/2)+iz\right)\\
        R_b &= \frac{1}{\sqrt{r^2+r^2\cot^2(\omega/2)}}\left(y+ix\right),
    \end{split}
\end{equation}
\noindent
which can be shown to be identically Eq. \ref{CK}.  
However, when implemented numerically, there exists a singularity at $\omega=0$ and $\omega=2\pi$, so we choose to implement Eq. \ref{CK} rather than treating $\omega=0$ as a separate case, and omitting $\omega=\pi$ altogether.  
Moreover, they used a recursive scheme to compute the $D$ matrices,

\begin{widetext}

\begin{equation}
    \begin{cases}
    D^j_{m m'} = \sqrt{\frac{j-m}{j-m'}}R_a^* D^{j-1/2}_{m+1/2, m'+1/2} - \sqrt{\frac{j+m}{j-m'}}R_b^* D^{j-1/2}_{m-1/2, m'+1/2}, &  m' \neq j \\
    D^j_{m m'} = \sqrt{\frac{j-m}{j+m'}}R_b D^{j-1/2}_{m+1/2, m'-1/2} + \sqrt{\frac{j+m}{j+m'}}R_a D^{j-1/2}_{m-1/2, m'-1/2}, & m' \neq -j
    \end{cases}
\end{equation}
\end{widetext}

Compared to the polynomial form Eq. \ref{horner}, the recursive form requires less floating point operations in general and is more efficient in serial calculations.  
However, in parallel architectures a polynomial form of the $D$-matrices is advantageous as no single term depends on another.  
During our implementation we found that using Numba's automatic parallelization\cite{numba} we were able to fuse all loops in the $D$-matrix calculation to achieve parallelization more so than algorithm when compared to the recursive version.  
This difference results in an improved scaling of the algorithm.
\bibliography{SNN-citations.bib}
\end{document}